\documentclass[12pt]{article}
\usepackage[english]{babel}
\usepackage[T1]{fontenc}
\usepackage{graphicx,subfigure}
\usepackage{amsmath,bm}
\usepackage[colorlinks=true,linkcolor=blue,citecolor=blue,urlcolor=blue]{hyperref}

\def\beq{\begin{equation}}
\def\eeq#1{\label{#1}\end{equation}}
\def\eeqn{\end{equation}}

\def\beqa{\begin{eqnarray}}
\def\eeqa#1{\label{#1}\end{eqnarray}}
\def\eeqan{\end{eqnarray}}

\let\bar=\overbar

\def\Dslash{\not{\hbox{\kern-4pt $D$}}}
\def\dslash{\not{\hbox{\kern-2pt $\del$}}}

\def\msb{{\bar{\ssstyle M \kern -1pt S}}}

\usepackage{fancyhdr,graphicx}
\def\Title#1{\begin{center} {\Large {\bf #1} } \end{center}}

\begin{document}

\Title{New insights on the hyperon puzzle from quantum Monte Carlo calculations}

\bigskip\bigskip


\begin{raggedright}

{\it 
F. Pederiva$^{\,1,2}$~~F. Catalano$^{\,3}$~~D. Lonardoni$^{\,4}$~~A. Lovato$^{\,4}$~~and S. Gandolfi$^{\,5}$ \\
\bigskip
$^{1}$Department of Physics,
University of Trento,
Via Sommarive 14,
I-38123 Trento,
Italy\\
\bigskip
$^{2}$INFN-Trento Institute for Fundamental Physics and Application,
Trento,
Italy\\
\bigskip
$^{3}$ Department of Physics and Astronomy,
\AA ngstr\"omlaboratoriet, L\"agerhyddsv\"agen,
75120 Uppsala,
Sweden\\
\bigskip
$^{4}$ Physics Division, Argonne National Laboratory, 
Lemont, Illinois 60439, USA\\
\bigskip
$^{5}$ Theoretical Division, Los Alamos National Laboratory, 
Los Alamos, New Mexico 87545, USA\\
}

\end{raggedright}

\section{Introduction}

The presence of hyperons in the inner core of a neutron star (NS) is still the subject
of an open debate. The mechanism promoting the formation of heavier, distinguishable
particles within dense interacting matter is quite intuitive at the non-relativistic
level, and relies essentially on the Pauli principle. However, the energy and pressure
reduction due to the presence of hyperons in the medium seems to be incompatible with the most
recent astronomical observations. In particular, most of non-relativistic models predict a softening
of the equation of state (EOS) which is too large to sustain a neutron star of mass $\sim 2M_\odot$ 
as the ones that have been recently observed~\cite{Demorest:2010,Antoniadis:2013}. 
This apparent inconsistency between NS mass observations and theoretical calculations 
is a long standing problem known as \emph{hyperon puzzle}.

At present it is still too difficult to derive a first-principle effective Hamiltonian directly from 
LQCD results in matter with strangeness. Therefore it is still necessary to rely on 
the scarce experimental data to work out a realistic interaction between hyperons
and nucleons in matter, and provide a reliable prediction for the EOS. 
We have been pursuing for a few years a program aiming to build up a phenomenological,
realistic potential along the tracks of the Argonne/Illinois model. This work was started
originally by Bodmer et al.~\cite{Bodmer:1984}, and continued in a series of papers by using
variational Monte Carlo algorithms
~\cite{Bodmer:1985,Bodmer:1988,Usmani:1995,Usmani:1995_3B,Usmani:1999,Usmani:2008,Imran:2014}. 
Our additional ingredient is the consistent use of auxiliary field diffusion Monte Carlo (AFDMC)
~\cite{Schmidt:1999,Carlson:2014} calculations
in oder to first assess the parameters of the interaction from measured hyperon separation
energies, and then to extrapolate the results to the case of neutron matter with strangeness.
Other strange degrees of freedom might be relevant to the end of a correct determination
of the EOS. However, in our approach we rely only on available
experimental data. Essentially no measurements are available for $\Sigma$ and $\Xi$ hypernuclei,
and therefore we exclude at the moment these channels from our treatment. The same 
argument applies to two- and many-hyperon forces, which, at present, are 
substantially unknown from the experimental point of view. 

Most of our results have been recently published~\cite{Lonardoni:2013,Lonardoni:2014,Lonardoni:2015}, 
and we refer to the original papers
for all the detailed aspects of the algorithm and the actual calculations.
In this proceeding we will briefly comment
on a particular aspect of the hyperon-nucleon-nucleon force that nicely illustrates
the difficulty in extracting the information on the Hamiltonian from experimental $\Lambda$ separation energies.

\section{A phenomenological, realistic  hyperon-nucleon interaction}

Within our non-relativistic many-body approach, $\Lambda$~hypernuclei and 
$\Lambda$-neutron matter are described in terms of pointlike nucleons and lambdas, 
with masses $m_N$ and $m_\Lambda$, respectively, whose dynamics are dictated by
the Hamiltonian:
\begin{align}
	H_{\rm nuc}	&=T_N+V_{NN}=\displaystyle\sum_{i}\frac{p_i^2}{2m_N}+\sum_{i<j}v_{ij}+\sum_{i<j<k}v_{ijk}\;,\\[1em]
	H_{\rm hyp}	&=H_{\rm nuc}+T_\Lambda+V_{\Lambda N}+V_{\Lambda NN}\nonumber\\[0.5em]
				&=\displaystyle H_{\rm nuc}+\sum_{\lambda}\frac{p_\lambda^2}{2m_\Lambda}+\sum_{\lambda i}v_{\lambda i}
				+\!\!\sum_{\lambda,i<j}v_{\lambda ij}\;,
\end{align}
where $A$ is the total number of baryons $A=\mathcal
N_N+\mathcal N_\Lambda$, latin indices $i,j=1,\ldots,\mathcal N_N$ label
nucleons, and the greek symbol $\lambda=1,\ldots,\mathcal N_\Lambda$
is used for $\Lambda$~particles. The nuclear potential includes
two- and three-nucleon contributions while in the strange sector we adopt explicit
$\Lambda N$ and $\Lambda NN$ interactions.

In the non-strange sector we employ a simplified interaction in order to make
the calculations feasible also for heavier hypernuclei. In particular we use
the sum of the Argonne AV4' interaction~\cite{Wiringa:2002}, plus the central repulsive term of the three-body
Urbana IX potential~\cite{Pudliner:1997}. This choice provides a realistic description of energies, densities
and radii. In Tab.~\ref{tab} we report some examples of the binding energies obtained 
from AFDMC calculations for a set of closed shell nuclei. 
\begin{table}
\centering{}%
\begin{tabular}{|c|c|c|c|c|}
\hline 
 & AV4' &  AV4'+UIX$_c$  & exp. & diff. (\%) 
\tabularnewline
\hline 
$^4$He  & -32.83(5)  & -26.63(3)  & -28.296 & 6  \tabularnewline
\hline 
$^{16}$O  & -180.1(4)  & -119.9(2)  & -127.619 & 6 \tabularnewline
\hline 
$^{40}$Ca  & -597(3)  & -382.9(6)  & -342.051 & 12 \tabularnewline
\hline 
$^{48}$Ca  & -645(3)  & -414.2(6)  & -416.001 & 0.5  \tabularnewline
\hline
\end{tabular}\medskip{}
 \caption{Energies of a few nuclei computed by AFDMC using the
 AV4' nucleon-nucleon potential, and including a three-body repulsive
 contribution taken from the UIX potential.  
\label{tab}}
\end{table}

In the strange sector the $\Lambda N$ interaction has been modeled with an Urbana-type
potential~\cite{Lagaris:1981}, consistent with the available $\Lambda p$
scattering data
\begin{align}
        v_{\lambda i}=v_{0}(r_{\lambda i})+\frac{1}{4}v_\sigma T^2_\pi(r_{\lambda i})\,{\bm\sigma}_\lambda\cdot{\bm\sigma}_i \;,
        \label{eq:V_LN}
\end{align}
where $v_0(r)=v_c(r)-\bar v\,T_{\pi}^{2}(r)$ is a central term. The terms
$\bar v=(v_s+3v_t)/4$ and $v_\sigma=v_s-v_t$ are the spin-average and
spin-dependent strengths, where $v_s$ and $v_t$ denote singlet- and
triplet-state strengths, respectively.  Note that both the spin-dependent
and the central radial terms contain the usual regularized one pion exchange tensor
operator $T_\pi(r)$
\begin{align}
        T_{\pi}(r)=\left[1+ \frac{3}{\mu_\pi r}+ \frac{3}{(\mu_\pi r)^2} \right]
        \frac{e^{-\mu_\pi r}}{\mu_\pi r}\Bigl(1-e^{-cr^2}\Bigr)^2\;,\label{eq:T_pi}
\end{align}
where $\mu_\pi$ is the reduced pion mass
\begin{align}
        \mu_\pi=\frac{m_{\pi^0}+2\,m_{\pi^\pm}}{3}\quad\quad\frac{1}{\mu_\pi}\simeq 1.4~{\rm fm} \;.
\end{align}
All the parameters defining the $\Lambda N$ potential can be found, for example, in
Ref.~\cite{Usmani:2008}.
The three-body potential $v_{\lambda ij}$ can be conveniently
decomposed in the 2$\pi$-exchange contributions
$v^{{2\pi}}_{\lambda ij}=v^{2\pi,P}_{\lambda ij}+v^{2\pi,S}_{\lambda ij}$ and a spin-dependent dispersive term 
$v_{\lambda ij}^{D}$  as follows:
\begin{align}
   v_{\lambda ij}^{2\pi,P}&=-\frac{C_P}{6}
           \Bigl\{X_{i\lambda}\,,X_{\lambda j}\Bigr\}\,{\bm\tau}_{i}\cdot{\bm\tau}_{j}\;,\label{eq:V_LNN_P}\\[0.7em]
   v_{\lambda ij}^{2\pi,S}&=
           C_S\,Z\left(r_{\lambda i}\right)Z\left(r_{\lambda j}\right)\,
           {\bm\sigma}_{i}\cdot\hat{\bm r}_{i\lambda}\,
           {\bm\sigma}_{j}\cdot\hat{\bm r}_{j\lambda}\,{\bm\tau}_{i}\cdot{\bm\tau}_{j}\;,\label{eq:V_LNN_S}\\[0.7em]
   v_{\lambda ij}^{D}&=W_D\,
           T_{\pi}^{2}\left(r_{\lambda i}\right)T^{2}_{\pi}\left(r_{\lambda j}\right)
           \!\!\bigg[1+\frac{1}{6}{\bm\sigma}_\lambda\!\cdot\!\left({\bm\sigma}_{i}+{\bm\sigma}_{j}\right)\bigg]\;.\label{eq:V_LNN_D}
\end{align}
The function $T_\pi(r)$ is the same as in Eq.~(\ref{eq:T_pi}),
while the $X_{\lambda i}$ and $Z(r)$ are defined by
\begin{align}
\begin{array}{rl}
        X_{\lambda i}&=Y_{\pi}(r_{\lambda i})\;{\bm\sigma}_{\lambda}\cdot{\bm\sigma}_{i}+
        T_{\pi}(r_{\lambda i})\;S_{\lambda i}\;,\\[0.7em]
        Z(r)&=\displaystyle\frac{\mu_\pi r}{3} \Bigl[Y_\pi(r)-T_\pi(r)\Bigr]\;,
\end{array}
\end{align}
where
\begin{align}
        Y_\pi(r)=\frac{e^{-\mu_\pi r}}{\mu_\pi r}\Bigl(1-e^{-cr^2}\Bigr)
        \label{eq:Y_pi}
\end{align}
is the regularized Yukawa potential and $S_{\lambda i}$ is the
usual tensor operator.
The range of parameters $C_P$, $C_S$ and $W_D$ that have been used in our calculations
can be found in Ref. \cite{Lonardoni:2014}.

\begin{figure}[htb]
\centering
\includegraphics[width=0.8\textwidth]{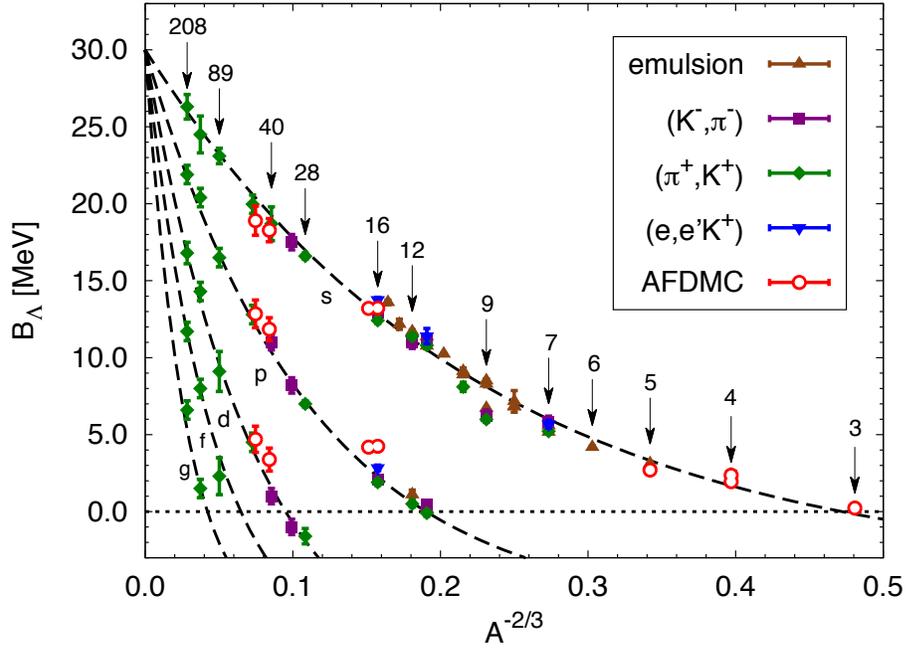}
\caption{Measured~\cite{Juric:1973,Cantwell:1974,Bertini:1979,Pile:1991,
Hasegawa:1996,Takahashi:2001,Hotchi:2001,Miyoshi:2003,Yuan:2006,Hashimoto:2006,
Cusanno:2009,Agnello:2010,Nakazawa:2010,Agnello:2011,Agnello:2012,Feliciello:2013,
Ahn:2013,Nakamura:2013,Tang:2014} and computed $\Lambda$~separation energies as a function of $A^{-2/3}$. 
Results for the $\Lambda$~particle in different single particle states are also shown.}
\label{fig:BL}
\end{figure}

AFDMC predictions for the $\Lambda$ separation energy $B_\Lambda$, defined as the difference in 
the binding energy of the core nucleus and the corresponding hypernucleus, are in good agreement 
with experimental data for hypernuclei up to $^{49}_{~\Lambda}$Ca and also for the $\Lambda$~particle in different 
single particle states (see Fig.~\ref{fig:BL}). 

\section{On the isospin dependence of the $\Lambda NN$ interaction}

As shown in Ref.~\cite{Lonardoni:2015}, the parametrization of the hyperon-nucleon potential
yielding the best prediction for $B_\Lambda$ provides a repulsion that is
large enough to prevent the appearance of strange degrees of freedom in the range of densities
found in the inner core of a NS. While this result might look as a possible
solution of the hyperon puzzle, a closer look to the interaction suggests
that the constraints coming from an analysis of the hypernuclear data
might not be sufficient to provide an accurate extrapolation to stellar
conditions. In the following we present a case study to effectively illustrates
this point.

The current version of the $\Lambda NN$ potential does not depend
on whether the two nucleons are in a singlet or a isospin triplet state.
For symmetric hypernuclei the Pauli principle suppresses any strong contribution 
from the $\Lambda nn$ or $\Lambda pp$ channels. On the other 
hand, in neutron matter or in matter at $\beta$-equilibrium the contribution
of the isospin triplet channel might become quite relevant. 
In order to test the sensitivity of our AFDMC
predictions for $B_\Lambda$ on the strength of the isospin triplet component, 
we consider the sum $v_{\lambda ij}^{T}$ of the $S$ and $P$ wave $2\pi$ exchange
terms:
\begin{align}
v_{\lambda ij}^{T}\,\bm\tau_i\cdot\bm\tau_j= \left[-\frac{C_P}{6}
                \Bigl\{X_{i\lambda}\,,X_{\lambda j}\Bigr\}+
C_S\,Z\left(r_{\lambda i}\right)Z\left(r_{\lambda j}\right)\,
                {\bm\sigma}_{i}\cdot\hat{\bm r}_{i\lambda}\,
                {\bm\sigma}_{j}\cdot\hat{\bm r}_{j\lambda}\right]{\bm\tau}_{i}\cdot{\bm\tau}_{j}\;.
\end{align}
The operator ${\bm\tau}_{i}\cdot{\bm\tau}_{j}$
can be written in terms of the projectors on the triplet $(T=1)$ and singlet $(T=0)$ nucleon isospin channels:
\begin{align}
	{\bm\tau}_{i}\cdot{\bm\tau}_{j}=-3\,P_{ij}^{T=0}+P_{ij}^{T=1}\;.
\end{align}
The potential can be then rewritten as:
\begin{align}
	v_{\lambda ij}^{T}\,\bm\tau_i\cdot\bm\tau_j=-3\,v_{\lambda ij}^{T}\,P_{ij}^{T=0}+C_T\,v_{\lambda ij}^{T}\,P_{ij}^{T=1}\;,
\end{align}
where the additional parameter $C_T$ gauges the strength and the sign of the isospin triplet contribution.
For $C_T=1$ the original parametrization of the three-body force is recovered.

\begin{figure}[htb]
\centering
\includegraphics[width=0.8\textwidth]{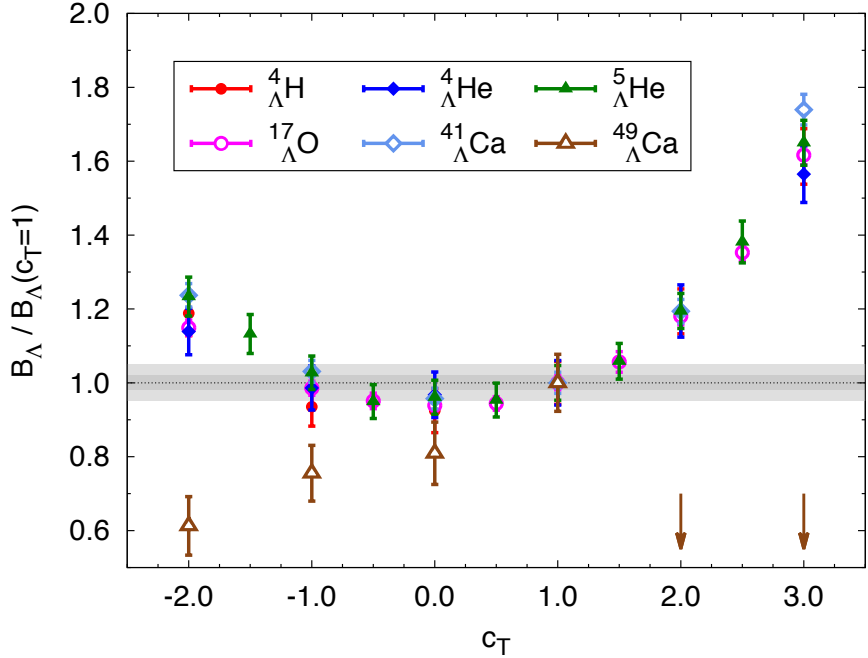}
\caption{$\Lambda$~separation energies normalized with respect to the $C_T=1$ case as a function of $C_T$.
	Grey bands represent the 2\% and 5\% variations of the ratio $B_\Lambda/B_\Lambda(C_T=1)$.
	Brown vertical arrows indicate the results for $^{49}_{~\Lambda}$Ca in the case of $C_T=2$ and $C_T=3$,
	outside the scale of the plot.}
\label{fig:CT}
\end{figure}

We have performed calculations for several hypernuclei ($^4_\Lambda$H, $^4_\Lambda$He, $^5_\Lambda$He, $^{17}_{~\Lambda}$O,
$^{41}_{~\Lambda}$Ca, $^{49}_{~\Lambda}$Ca) varying the $C_T$ parameter in the range $[-2;3]$. 
The results are reported in Fig.~\ref{fig:CT} as the ratio of $B_\Lambda(C_T)$ and $B_\Lambda$ at $C_T=1$, 
i.e. with the original parametrization.

It can be observed that the sensitivity of the results on the value of $C_T$ is not large over the whole
interval $-1.0\le C_T \le 1.5$, while strong deviations appear beyond this range. In general, outside of this
range $B_\Lambda$ tends to increase with respect to the original value. The only is exception is
$^{49}_{~\Lambda}$Ca, where the sensitivity appears to be larger, and $B_\Lambda$ tends to decrease with respect
to the original value. Given the substantial asymmetry of this hypernucleus, we can infer that the isospin triplet
channel could be properly constrained only by looking at the binding energy of strongly asymmetric
hypernuclei. Extrapolating results to neutron matter without taking into account this feature of the
interaction might in principle lead to misleading results. 

\section{Conclusions}
In the last years auxiliary field diffusion Monte Carlo has been used  to 
assess the properties of hypernuclear systems, from light- to medium-heavy
hypernuclei and hyper-neutron matter. One of the main findings is the key
role played by the three-body hyperon-nucleon-nucleon interaction 
in the determination of the hyperon separation energy of hypernuclei and 
as a possible solution to the hyperon puzzle.
However, there are still aspects of the employed hypernuclear potential that 
remain to be carefully investigated. For instance, we showed that the isospin
dependence of the $\Lambda NN$ force, which is crucial in determining the 
NS structure, is poorly constrained by the available experimental data. 
Future experiments on highly asymmetric hypernuclei such as $^{49}_{~\Lambda}$Ca, 
$^{91}_{~\Lambda}$Zr or even $^{209}_{~~\Lambda}$Pb would 
pin down fundamental properties of the hyperon-nucleon forces. This would thereby allow
for a substantial step forward in understanding the deep connections between the physics at the km scale typical of NSs 
and the properties of matter at the fm scale that can be efficiently explored in terrestrial experiments.

\subsection*{Acknowledgement}

This work was supported by the U.S. Department of Energy, 
Office of Science, Office of Nuclear Physics, under the NUCLEI SciDAC grant
(D.L., A.L., S.G.), by the Department of Energy, Office of Science, Office of Nuclear Physics, 
under Contract No. DE-AC02-06CH11357 (A.L.), and by DOE under Contract No. DE-AC02-05CH11231 
and Los Alamos LDRD grant (S.G.).
This research used resources of the National
Energy Research Scientific Computing Center (NERSC), which is supported by
the Office of Science of the U.S. Department of Energy under Contract No.
DE-AC02-05CH11231.



\begin{thebibliography}{99}
	
\bibitem{Demorest:2010} 
 P. B. Demorest, T. Pennucci, S. M. Ransom, M. S. E.
Roberts, and J. W. T. Hessels, Nature 467, 1081 (2010).
\bibitem{Antoniadis:2013}
 J. Antoniadis et al., Science 340, 1233232 (2013).
\bibitem{Bodmer:1984}
 A. R. Bodmer, Q. N. Usmani, and J. Carlson, Phys. Rev.
C 29, 684 (1984).
\bibitem{Bodmer:1985}
 A. R. Bodmer and Q. N. Usmani, Phys. Rev. C 31, 1400
(1985).
 \bibitem{Bodmer:1988}
 A. R. Bodmer and Q. N. Usmani, Nucl. Phys. A 477,
621 (1988).
\bibitem{Usmani:1995}
 A. A. Usmani, Phys. Rev. C 52, 1773 (1995).
\bibitem{Usmani:1995_3B}  
 A. A. Usmani, S. C. Pieper, and Q. N. Usmani, Phys.
Rev. C 51, 2347 (1995).
\bibitem{Usmani:1999}
 Q. N. Usmani and A. R. Bodmer, Phys. Rev. C 60,
055215 (1999).
\bibitem{Usmani:2008}
 A. A. Usmani and F. C. Khanna, J. Phys. G 35, 025105
(2008).
\bibitem{Imran:2014}
 M. Imran, A. A. Usmani, M. Ikram, Z. Hasan, and F. C.
Khanna, J. Phys. G 41, 065101 (2014).
\bibitem{Wiringa:2002}
 R. B. Wiringa, S. C. and Pieper,
Phys. Rev. Lett. 89, 182501 (2002). 
\bibitem{Pudliner:1997}
 B. S. Pudliner, V. R. Pandharipande, J. Carlson, S. C. Pieper, and R. B. Wiringa,
Phys. Rev. C 56, 1720 (1997).
\bibitem{Schmidt:1999}
K. E. Schmidt and S. Fantoni, Phys. Lett. B 446, 99
(1999).
\bibitem{Carlson:2014}
 J. Carlson, S. Gandolfi, F. Pederiva, S. C. Pieper, R. Schiavilla, et al.
arXiv:1412.3081.
\bibitem{Lonardoni:2013}
 D. Lonardoni, S. Gandolfi, and F. Pederiva, Phys. Rev.
C 87, 041303 (2013).
\bibitem{Lonardoni:2014}
 D. Lonardoni, F. Pederiva, and S. Gandolfi, Phys. Rev.
C 89, 014314 (2014).
\bibitem{Lonardoni:2015}
 D. Lonardoni, A. Lovato, S. Gandolfi, and F. Pederiva,
 Phys. Rev. Lett. 114, 092301 (2015).
\bibitem{Lagaris:1981}
 I. Lagaris and V. Pandharipande, Nucl. Phys. A 359, 331
(1981).
\bibitem{Juric:1973}
 M. Juri\v{c} et al., Nucl. Phys. B 52, 1-30 (1973).
\bibitem{Cantwell:1974}
 T. Cantwell et al., Nucl. Phys. A 236, 445 (1974).
\bibitem{Bertini:1979}
 R. Bertini et al., Phys. Lett. B 83, 306 (1979).
\bibitem{Pile:1991}
 P. Pile et al., Phys. Rev. Lett. 66, 2585 (1991).
\bibitem{Hasegawa:1996}
 T. Hasegawa et al., Phys. Rev. C 53, 1210 (1996).
\bibitem{Takahashi:2001}
 H. Takahashi et al., Phys. Rev. Lett. 87, 212502 (2001).
\bibitem{Hotchi:2001}
 H. Hotchi et al., Phys. Rev. C 64, 044302 (2001).
\bibitem{Miyoshi:2003}
 T. Miyoshi et al., Phys. Rev, Lett. 90, 232502 (2003).
\bibitem{Yuan:2006}
 L. Yuan et al., Phys. Rev. C 73, 044607 (2006).
\bibitem{Hashimoto:2006}
 O. Hashimoto et al., Progr. Part. Nucl. Phys. 57, 564 (2006).
\bibitem{Cusanno:2009}
 F. Cusanno et al., Phys. Rev. Lett. 103, 202501 (2009).
\bibitem{Agnello:2010}
 M. Agnello et al., Nucl. Phys. A 835, 414 (2010).
\bibitem{Nakazawa:2010}
 K. Nakazawa, Nucl. Phys. A 835, 207 (2010).
\bibitem{Agnello:2011}
 M. Agnello et al., Phys. Lett. B 698, 219 (2011).
\bibitem{Agnello:2012}
 M. Agnello et al., Phys. Rev. Lett. 108, 042501 (2012).
\bibitem{Feliciello:2013}
 A. Feliciello, Mod. Phys. Lett. A 28, 1330029 (2013).
\bibitem{Ahn:2013}
 J. K. Ahn et al., Phys. Rev. C 88, 014003 (2013).
\bibitem{Nakamura:2013}
 S. Nakamura et al., Phys. Rev. Lett. 110, 012502 (2013).
\bibitem{Tang:2014}
 L. Tang et al., Phys. Rev. C 90, 034320 (2014).

\end{thebibliography}
\end{document}